\documentclass[twocolumn,showpacs,preprintnumbers,amsmath,amssymb,pra]{revtex4}
\usepackage{graphicx}
\usepackage{dcolumn}
\usepackage{bm}

\begin{document}

\title{Phase diagrams of 2D and 3D disordered Bose gases in the local density approximation}
\author{Thomas Bourdel
\footnote{Email: thomas.bourdel@institutoptique.fr}
}
\affiliation{Laboratoire Charles Fabry, UMR 8501, Institut d'Optique, CNRS, Univ Paris Sud 11, Avenue Augustin Fresnel, 91127 PALAISEAU CEDEX}
\date{\today}
\begin{abstract}
We study the superfluid transitions in bidimensional (2D) and tridimensional (3D) disordered and interacting Bose gases. We work in the limit of long-range correlated disorder such that it can be treated in the local density approximation. We present the superfluid transition curves both in the disorder-temperature plane well as in the disorder-entropy plane in 2D and 3D Bose gases. Surprisingly, we find that a small amount of disorder is always favorable to the apparition of a superfluid. Our results offer a quantitative comparison with recent experiments in 2D disordered ultra-cold gases, for which no exact theory exists.
\end{abstract}


\maketitle

\section{Introduction}

Disorder is an essential ingredient for the understanding of transport in condensed matter physics. At low temperature, it affects the conductivity of a metal and it even induces phase transitions to insulating states \cite{Imada1998}. Superconductor-insulator transitions \cite{Gantmakher2010} are observed in thin metal films \cite{Goldman1998} and, in high-Tc superconductors, where doping intrinsically introduces inhomogeneities in the CuO-planes \cite{Pan2001}. Understanding the complex interplay between disorder and interactions in these systems remains a major challenge. Whereas the above mentioned electronic systems are fermionic, superconductivity originates from the bosonic nature of Cooper pairs. As long as disorder does not break the Cooper pairs, the problem is reduced to a study of dirty bosons \cite{Fisher1989, Bollinger2011}. It has mainly been studied numerically in the framework of the disordered Bose-Hubbard model \cite{Fisher1989, Krauth1991}. This model was realized experimentally in 1D and 3D ultra-cold gases in disordered lattices \cite{Fallani2007, White2009, Pasienski2010}. The physics is complicated by the presence of a Mott insulating transition, which occurs in the absence of disorder. 

In this paper, we focus on a different model with correlated disorder in a continuous Bose system. This model has been studied previously in 3D \cite{Pilati2010}.  Our work is motivated by recent experiments studying the superfluid transition in disordered 2D ultra-cold atomic Bose gases \cite{Allard2012, Beeler2012}. A schematic phase diagram for such a system has been proposed and at low temperatures and low disorder, a superfluid is expected \cite{Aleiner2009}. However, the exact shape of the superfluid region is not known. In this paper, we calculate the superfluid transition curve for a disorder that is correlated at long range such that we can make use of the local density approximation. In this limit, we find quantitative predictions in both 2D and 3D. In 3D, we rely on a mean-field description of the Bose gas \cite{Huang1987}  whereas in 2D, we use the equation of state from Monte-Carlo-simulations \cite{Prokofev2002}. Interestingly, we find a very different quantitative behavior in 2D and 3D and that a small amount of disorder (either added reversibly or at constant temperature) is always favorable to the appearance of a superfluid.

The paper is structured as follows. In the second section, we define our system parameters and the conditions of validity of the local density approximation. In the following section, we calculate the critical disorder at zero temperature in homogeneous and harmonically trapped Bose gases in 3D and 2D. In the fourth section, the superfluid transition curves are found as a function of disorder and temperature. In the fifth section, we extend our results to get the curves as a function of entropy which is more convenient for the comparison to experiments which is presented in the sixth section.

\section{The local density approximation in a disordered potential}

We consider Bose gases with repulsive contact interaction, i.e. with a positive scattering length $a$, such that there is no collapse of the wave function. 
The 2D gases are supposed to be harmonically confined in the vertical $z$ direction with a frequency $\omega_z/2 \pi$ such that the atoms occupies in this direction a single quantum state, whose size is $l_z=\sqrt{\hbar/m \omega_z}$, where $\hbar$ is the Planck constant and $m$ the atomic mass. We also suppose that $a \ll l_z$, such that the interactions can be described by 3D scattering.  The interaction parameter is then $g_\textrm{2D}=(\hbar^2/m) \tilde{g}=(\hbar^2/m) \sqrt{8 \pi} a/l_z$ \cite{Hadzibabic2011}. In three dimensions, the interaction parameter is $g_\textrm{3D}=4 \pi \hbar^2 a/m$. The healing length of a gas of density $n_\textrm{d}$, where $\textrm{d}$ stands for the dimension, is $\xi=\hbar/\sqrt{m g_\textrm{dD} n_\textrm{d}}$. It reflects the size over which the density of a condensate (or quasi-condensate in 2D) comes back to its original value when a local perturbation is imposed. 

We consider a disorder created by a laser speckle, with a blue detuned frequency as compared to the atomic transition frequency. In this case, the optical potential $V({\bf r})$ is repulsive \cite{Grimm2000}. We suppose that the speckle intensity has a gaussian correlation function whose $1/\sqrt{e}$ radius is $\sigma$. The mean disorder potential $\bar{V}$ is proportional to the laser power. Knowing $\sigma$ and $\bar{V}$, all statistical properties of the disorder are then known \cite{Clement2006}. In particular, the probability density of having a disorder $V$ is an exponential $P(V)=\exp (-V/\bar{V})/\bar{V}$. The classical percolation threshold for such a disorder is proportional to the mean disorder $\bar{V}$ and we call $\alpha$ the proportionality coefficient. In 2D,  $\alpha_\textrm{2D}=$0.52 \cite{Smith1979, Weinrib1982, Pezze2011} and, in 3D, $\alpha_\textrm{3D} \sim 4\,10^{-4}$ \cite{Pilati2010}. The much lower value of $\alpha$ in 3D as compared to 2D reflects the fact that it is easier to find ways around the peaks of the disorder in 3D.

The local density approximation (LDA) assumes that the gas can be considered locally as homogenous with a local chemical potential $\mu_\textrm{local}=\mu-V_\textrm{ext}$, where $V_\textrm{ext}$ is the external potential, i.e. the sum of the trapping and disorder potential. In typical experimental conditions in non-disordered harmonic traps, the LDA tends to give extremely accurate results. In order for this approximation to be valid in a disorder potential, $\sigma$ must be much larger than all characteristic lengths of the gas such as the healing length $\xi$ and the thermal De Broglie wavelength $\lambda_\textrm{dB}=\sqrt{2 \pi \hbar^2/m k_\textrm{B} T}$, where $k_\textrm{B}$ is the Boltzmann factor and $T$ the temperature.

\section{Superfluid to insulator quantum phase transition}

In 2D and 3D, homogeneous interacting  Bose gases at zero temperature are always superfluid independently of the density. Within the local density approximation, a superflow in the presence of disorder can then exist,  if the region of non-zero density percolates through the sample. The quantum phase transition ($T=0$) is thus reduced to the percolation of the density. 
The density of the gas $n_\textrm{d}$ can be found from the Thomas-Fermi approximation of the Gross-Pitaevskii equation \cite{Sanchez-Palencia2006} which corresponds to the local density approximation for a {(quasi-)condensate} $n_\textrm{d}=\textrm{Max}(\mu_\textrm{local}, 0)/g_\textrm{dD}$. The gas will thus be superfluid when the region where $\mu_\textrm{local}>0$ percolates, i.e. when the global chemical potential $\mu$ exceeds the classical percolation threshold of the disorder  $\alpha \bar{V}$. For $0<\mu<\alpha_\textrm{3D} \bar{V}$, local superfluid islands exist in the disorder minima but it does not lead to global superfluidity.

In a disordered Bose gas at zero temperature in a box the number of atoms is\,:
\begin{equation}
N=\int \textrm{Max}(\mu-V({\bf r}), 0)/g_\textrm{dD} \textrm{d}^dr.
\end{equation}
When averaging of the disorder, we can replace the position dependance of the disorder by an integral\,:
\begin{equation}
\langle N \rangle=\int \int \textrm{Max}(\mu-V, 0)/g_\textrm{dD} \textrm{d}^dr P(V) \textrm{d}V.
\end{equation}
The integration over the volume is then trivial. At the transition point, i.e. for $\mu=\alpha \bar{V}$, we find the critical mean density
\begin{equation}
\langle n_\textrm{dD}^\textrm{cond} \rangle= \frac{\bar{V}}{g_\textrm{dD}} (-1+\exp(-\alpha_\textrm{dD})+\alpha_\textrm{dD}) \underset{{\alpha_\textrm{dD} \rightarrow 0}}{\approx} \frac{\bar{V}}{2 g_\textrm{dD}} \alpha_\textrm{dD}^2 \textrm{ ,}
\end{equation}
which is valid in both 2D and 3D. The expansion for low value of $\alpha_\textrm{dD}$ is especially accurate in 3D. In 2D, we find the critical disorder strength $\bar{V}_\textrm{crit} \approx 8.7 g_\textrm{2D} \langle n_\textrm{2D}^\textrm{cond} \rangle$, whereas in 3D, $\bar{V}_\textrm{crit} \approx 1.2\,10^7 g_\textrm{3D} \langle n_\textrm{3D}^\textrm{cond} \rangle$. A much higher disorder is thus needed in order to reach the superfluid quantum phase transition in 3D.

In the presence of an additional harmonic confinement, we suppose that the gas is superfluid as soon as the superfluid threshold is obtained in the center of the trap where the density is maximal. The calculation can then be done along the same lines with the trapping potential entering in the expression of the local chemical potential. The spacial integration leads to different results in 2D and 3D for the critical atom number as a function of the disorder\,:
\begin{gather}
\langle N_\textrm{2D}^\textrm{cond}\rangle=\pi \frac{\bar{V}^2}{\tilde{g}\hbar^2 \omega^2} (2-2\exp(-\alpha_\textrm{2D})-2\alpha_\textrm{2D}+\alpha_\textrm{2D}^2)
\\
\langle N_\textrm{3D}^\textrm{cond}\rangle=\frac{4 \sqrt{2}}{15}  \frac{\bar{V}^{5/2}}{(\hbar \omega)^{5/2} a \sqrt{m \omega /\hbar}} \int_0^{\alpha_\textrm{3D}} (\alpha_\textrm{3D}-v)^{5/2}\textrm{e}^{-v} \textrm{d}v 
\\
\underset{{\alpha_\textrm{3D} \rightarrow 0}}{\approx}\frac{4 \sqrt{2}}{15}  \frac{\bar{V}^{5/2}}{(\hbar \omega)^{5/2} a \sqrt{m \omega /\hbar}} \frac{2}{7}\alpha_\textrm{3D}^{7/2} \nonumber
\textrm{ .}
\end{gather}
Here, $\omega$ is the geometrical mean of the two in-plane trap oscillation frequencies in 2D and of the three oscillation frequencies in 3D. 

\section{Phase diagram of 2D and 3D Bose gases as a function of temperature and disorder}

We now look for the temperature dependance of the superfluid transition in the presence of disorder. In the previous section, we have found the critical disorder at zero temperature. The critical temperature with no disorder is known both in 2D \cite{Prokofev2002} and 3D \cite{Huang1987}.  In the following, we study the shape of the superfluid transition curve which connects these two points in the temperature-disorder phase diagram. We separate the 3D case, in which a mean field approximation is possible, from the 2D case, in which one has to rely on classical field numerical calculations.

\subsection{The superfluid transition at finite temperature in a 3D gas}

A 3D Bose gas at low temperatures is well described by two components, a condensate fraction and a thermal fraction. For typical interaction parameters used in experiments, the interaction between the atoms in the thermal cloud can be neglected and the interaction between the thermal cloud and the condensate can be treated in a mean field approach. Similarly to the $T=0$ case, the disordered 3D Bose gas is superfluid if the condensate density percolates through the sample. The critical chemical potential is thus equal to $\alpha_\textrm{3D} \bar{V}$. In order to know when the phase transition is reached, we have to calculate the relation between the atom number, the temperature and the chemical potential in the presence of disorder. 

The thermal cloud can be treated in the semi-classical approximation. The atom number for a Bose gas in a box is \cite{Huang1987} 
\begin{gather}
N_\textrm{th}=
\int \frac{\textrm{d}^3k \textrm{d}^3r}{(2 \pi)^3} \frac{1}{ \textrm{e}^{\beta (\hbar^2 k^2/2m+2g_\textrm{3D} n^\textrm{cond}_\textrm{3D}({\bf r})+V({\bf r})-\mu)}-1} \textrm{ ,}
\end{gather}
where $\beta=1/k_\textrm{B}T$ and 
where $2 g_\textrm{3D} n^\textrm{cond}_\textrm{3D}$ is the term due to the interaction with the condensate part in the mean field approximation. After averaging over the disorder, and integrating over $r$ and $k$, the mean thermal density is 
\begin{gather}
\langle n_\textrm{th} \rangle \lambda_\textrm{dB}^3=\int_0^\infty \textrm{e}^{-v} \textrm{PolyLog}\left[\frac{3}{2}, \textrm{e}^{-|\beta \mu-\beta \bar{V}v)|}\right] \textrm{d}v\textrm{ ,}\\
\text{where } \textrm{PolyLog} \left[n, a \right]=\sum_{l=0}^\infty a^l/l^n \text{ ,}
\end{gather}
where the Polylog function is the result of the integration over $k$ and where the absolute value comes from the mean field interaction with the condensate.

The total mean density is then $\langle n_\textrm{tot}\rangle=\langle n_\textrm{3D}^\textrm{cond}\rangle+\langle n_\textrm{th}\rangle$, where $\langle n_\textrm{3D}^\textrm{cond}\rangle$ comes from equation 3 and $\langle n_\textrm{th}\rangle$ comes from equation 7. At the phase transition, the chemical potential is  $\mu=\alpha_\textrm{3D} \bar{V}$ and the conservation of the density gives the relation between the disorder strength and the critical temperature. It can be written in a dimensionless form using the quantities $V'=\bar{V}/\langle n_\textrm{tot}\rangle g_\textrm{3D}$, $T'=(\langle n_\textrm{tot}\rangle \lambda_\textrm{dB}^3)^{-2/3}$, and $a'=(\langle n_\textrm{tot}\rangle a^3)^{1/3}$, which quantifies the interaction strength. The equation relating $V'$ and $T'$ at the phase transition is then
\begin{gather}
1=V'((-1+\exp(-\alpha_\textrm{3D})+\alpha_\textrm{3D}))\\
+{T'}^{3/2}\int \textrm{e}^{-v} \textrm{PolyLog}\left[\frac{3}{2}, \textrm{e}^{-2 |\alpha_\textrm{3D}-v| a' V'/T'} \right] \textrm{d}v \text{ .}
\end{gather}
It can be solved numerically.

\begin{figure}[htbp!]
\includegraphics[width=0.48\textwidth]{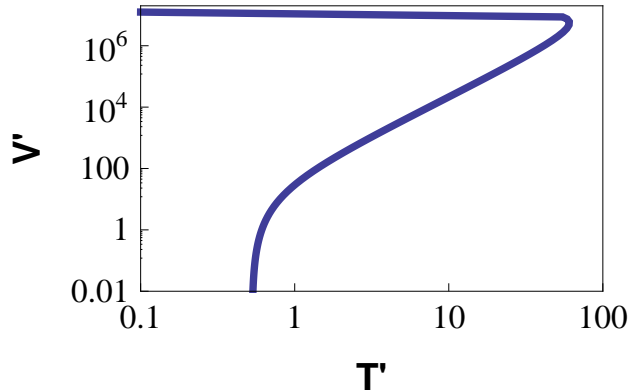}
\caption{Superfluid phase transition of a disordered 3D Bose gas in a box. We have fixed $\langle n_\textrm{tot}\rangle a^3=10^{-6}$. At low temperature and low disorder the gas is superfluid.}
\label{VT}
\end{figure}
The boundary between the normal and superfluid phase for a disordered Bose gas in a box is plotted in fig.  \ref{VT}. For low disorder, we find that the critical temperature increases with disorder. This is due to the compression of the gas by the disorder, which leads to an increase of the effective phase space density. When the disorder approaches its critical value at $T=0$ corresponding to $V'\approx 1.2\,10^7$, the increase of the percolation threshold starts to play a role and the critical temperature quickly goes to zero. 

A similar calculation, albeit slightly more technical, can be done in the presence of an harmonic confinement. In this case, natural non-dimensional parameters are $T'=k_\textrm{B}T/\hbar \omega N^{1/3}$, $V'=\bar{V}/\hbar \omega N^{1/3}$, and the strength of the interaction is characterized by $a'=N^{1/6}a \sqrt{m \omega /\hbar}$\,.The equation of the superfluid boundary is then
\begin{gather}
1=\frac{4 \sqrt{2}}{15a'}V'^{5/2} \int_0^\alpha (\alpha_\textrm{3D}-v)^{5/2}\textrm{e}^{-v} \textrm{d}v  \nonumber
 \\
+{T'}^{3/2}{V'}^{3/2}  \int_0^\infty (\frac{2}{\sqrt{\pi}}\sqrt{v}- \textrm{e}^{-v} \textrm{Erfi} \left[\sqrt{v}\right]) \nonumber
\\
\times \textrm{PolyLog} \left[ \frac{3}{2}, \exp(-|\alpha_\textrm{3D}-v|\frac{V'}{T'})\right] \textrm{d}v \text{ ,}
\end{gather}
where Erfi is the imaginary error function.

\begin{figure}[htbp!]
\includegraphics[width=0.48\textwidth]{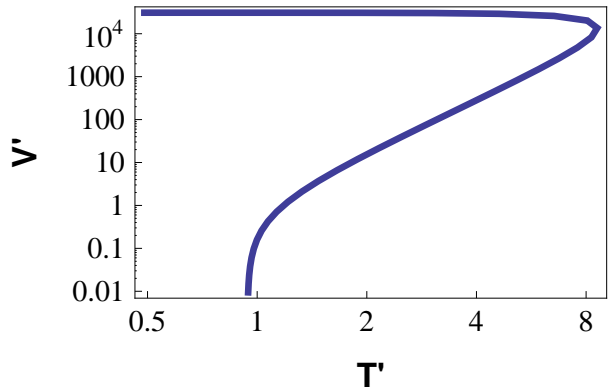}
\caption{Superfluid phase transition of a disordered 3D Bose gas in an harmonic trap. We have fixed $a'=N^{1/6}a \sqrt{m \omega /\hbar}=0.023$. At low temperature and low disorder the gas is superfluid.}
\label{VTharm}
\end{figure}
The boundary between the normal and superfluid phase for a disordered Bose gas in an harmonic trap is plotted in fig.  \ref{VTharm}. The interpretation is the same as before except that the disorder induced enhancement for the critical temperature is not as strong because heating the gas results in a further spreading in the trap and in a reduced density.

\subsection{The superfluid transition at finite temperature in a 2D gas}

The physics of 2D gases is more complex than the one of 3D gases because the effects of interactions are much stronger and cannot be captured in a mean field approximation.  The superfluid phase transition at finite temperature is of Berezinskii-Kosterlitz-Thouless type \cite{Berezinskii1972, Kosterlitz1973, Hadzibabic2011} and is intrinsically a beyond mean-field effect. As a consequence it is not possible to separate the gas into two components. The phase space density of the homogeneous 2D Bose gas $D=n_\textrm{2D} \lambda_\textrm{dB}^2$ has to be a function of $\mu/T$ and $\tilde{g}$ because the interaction parameter $\tilde{g}$ is dimensionless. This equation of state is not analytically known. It has been computed in Monte-Carlo simulations close to the phase transition \cite{Prokofev2002}. Further from the transition, simple analytic forms are available. At low phase-space density the mean-field approximation works, and at high phase space density the Thomas-Fermi limit is rapidly approached. Numerically, it is possible to interpolate between the different regimes with a good accuracy. 

From Monte-Carlo studies, we know that a finite superfluid fraction is expected at a critical value of the ratio $\mu_\textrm{crit}/k_\textrm{B}T \approx \tilde{g}\textrm{Log}(13.2/\tilde{g})/\pi$ which depends on the interaction parameter $\tilde{g}$ \cite{Prokof'ev2001}. In the local density approximation, we expect the gas to be superfluid if the regions of finite superfluid fraction percolate, i.e. if the difference of chemical potential $\mu-\mu_\textrm{crit}$ is larger than $\alpha \bar{V}$. To find the phase transition for a Bose gas in a box, the equation to solve is thus\,:
\begin{equation}
\langle n_\textrm{2D} \rangle \lambda_\textrm{dB}^2=\int P(V) D(\beta(\mu_\textrm{crit}+\alpha_\textrm{2D} \bar{V}-V), \tilde{g})\textrm{d}V \text{ .}
\end{equation}

\begin{figure}[htbp!]
\includegraphics[width=0.48\textwidth]{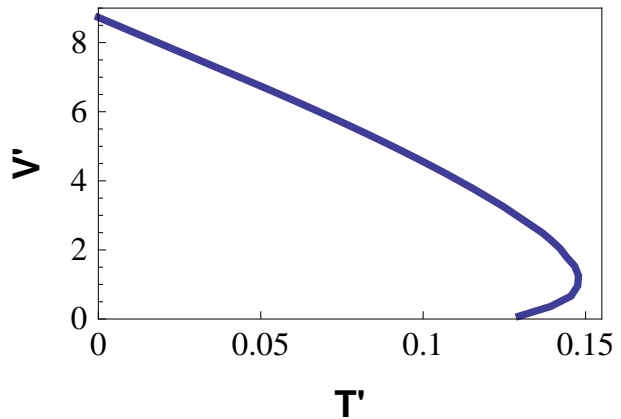}
\caption{Superfluid phase transition of a disordered 2D Bose gas in a box. $\tilde{g}=0.096$. At low temperature and low disorder the gas is superfluid.}
\label{VT2Dhom}
\end{figure}
Using the non dimensional parameters $V'=\bar{V}/g_\textrm{2D}\langle n_\textrm{2D} \rangle$, $T'=1/\langle n_\textrm{2D} \rangle \lambda_\textrm{dB}^2$, and $\tilde{g}$, the phase diagram takes the form shown in fig.\,\ref{VT2Dhom}. At $V=0$, the phase transition takes place at $\langle n_\textrm{2D} \rangle \lambda_\textrm{dB}^2 \approx 8 \approx 1/0.125$ as predicted in the absence of disorder for $\tilde{g}=0.96$ \cite{Prokofev2002}. A $T=0$, we recover the moderate critical value expected from the treatment using the Thomas-fermi approximation. In between, the shape of the phase diagram is different from the 3D case. We interpret this as a consequence of the much higher percolation threshold in 2D. In the phase diagram in 2D, there is no region where the percolation threshold is irrelevant in contrast to the 3D case. Nevertheless at small disorder, the effect of compression is still dominant and the critical temperature increases with disorder albeit by a small amount.

\begin{figure}[htbp!]
\includegraphics[width=0.48\textwidth]{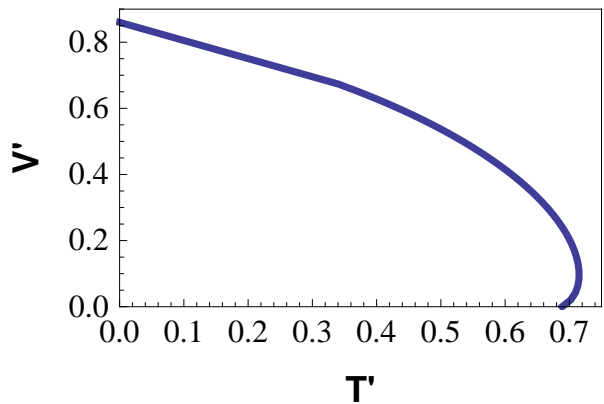}
\caption{Superfluid phase transition of a disordered 2D Bose gas in a harmonic trap. $\tilde{g}=0.096$. At low temperature and low disorder the gas is superfluid.}
\label{VT2Dharm}
\end{figure}
Again, a similar calculation can be made in an harmonic trap. In this case, we chose $V'=\bar{V}/\hbar \omega N^{1/2}$ and $T'=k_\textrm{B} T/\hbar \omega N^{1/2}$ in addition to $\tilde{g}$ as non-dimensional variables. The phase diagram is then shown in fig. \ref{VT2Dharm}. The initial increase of the critical temperature with disorder is still present although it is a relatively weak effect.

\section{Critical entropy as a function of disorder}

In all the previous cases, the addition of a small amount of disorder is favorable to the apparition of a superfluid fraction, in the sense that the critical temperature increases. This is due to the fact that the disorder compresses the gas in a smaller region and therefore leads locally to an increase of the phase space density (at constant temperature). In a recent experiment with ultra-cold atoms \cite{Allard2012}, the disorder is added reversibly after the preparation of the gas. In this case, the entropy is conserved (as well as the number of particles) but the temperature increases due to the effective compression of the gas. In the following, we derive the phase diagram as a function of disorder and entropy per particle, which is more directly related to the experiment. 

In the local density approximation, the calculation of the critical entropy can be done as soon as the entropy of a homogeneous gas is known. In the 3D case, as before the semi-classical approximation can be used and the entropy per unit of volume is \cite{Huang1987}
\begin{gather}
S=k_\textrm{B} \lambda_\textrm{dB}^{-3}{(}\frac{5}{2}\textrm{PolyLog}\left[\frac{5}{2}, \textrm{e}^{\mu_\textrm{local}/k_\textrm{B}T} \right] \nonumber 
\\
-\frac{\mu_\textrm{local}}{k_\textrm{B}T}\textrm{PolyLog}\left[\frac{3}{2}, \textrm{e}^{\mu_\textrm{local}/k_\textrm{B}T} \right]{)}.
\end{gather}

\begin{figure}[htbp!]
\includegraphics[width=0.48\textwidth]{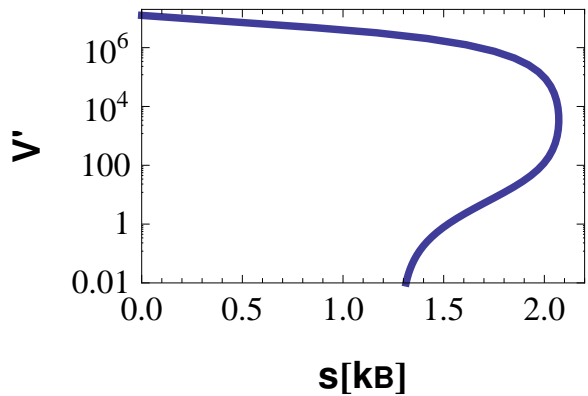}
\caption{Superfluid phase transition of a disordered 3D Bose gas in a box as a function of entropy per particle and disorder. $\langle n_\textrm{tot}\rangle a^3=10^{-6}$. At low entropy and low disorder the gas is superfluid.}
\label{VS3Dhom}
\end{figure}
This expression can be used to find the phase diagram as a function of the entropy and the disorder. It is plotted in fig. \ref{VS3Dhom}. Without disorder, the critical entropy per particle is the one that is found at the Bose-Einstein condensation threshold in a box. When increasing the disorder, the critical entropy first increases to a value close to 2.1. This is a  consequence of the change of the shape of the density of state at low energy \cite{Pinkse1997} from proportional to $\sqrt{E}$ without disorder to proportional to $E^{3/2}$ with disorder. Only at a disorder of the order of the critical disorder at $T=0$, the critical entropy decreases. In a 3D experiment, although it may appear counterintuitive, it is thus possible to induce superfluidity in slowly adding some disorder. A similar statement holds in the 3D harmonically trapped Bose gas, although we do not show the plot in this paper.

In the 2D case, the calculation of the entropy is possible as soon as the equation of state is known because of the scale invariance in 2D interacting Bose gases \cite{Yefsah2011}. More precisely, the entropy per particle is $S_\textrm{2D}=k_\textrm{B}(2P_\textrm{red}/D-\mu/k_\textrm{B}T)$, where the reduced pressure $P_\textrm{red}=\int_{-\infty}^{\mu/k_\textrm{B}T} D \textrm{d}(\mu/k_\textrm{B}T)$  can be calculated from the equation of state. 

\begin{figure}[htbp!]
\includegraphics[width=0.48\textwidth]{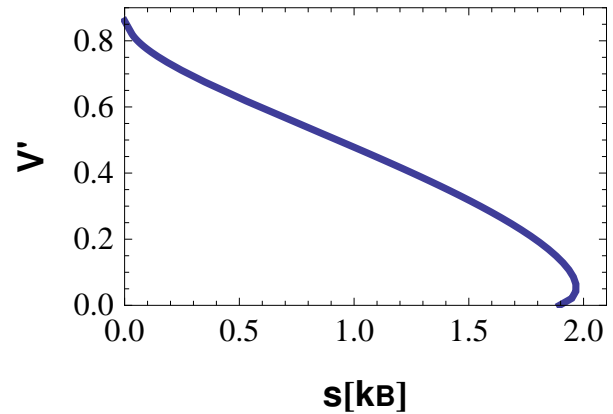}
\caption{Superfluid phase transition of a disordered 2D Bose gas in a harmonic trap as a function of entropy per particle and disorder. $\tilde{g}=0.096$. At low entropy and low disorder the gas is superfluid.}
\label{VS2Dharm}
\end{figure}
In fig. \ref{VS2Dharm}, the phase diagram of a harmonically trapped 2D gas is plotted as a function of entropy per particle and disorder. For small disorder, the critical entropy still increases with disorder and it is thus favorable for the appearance of a superfluid fraction. However, this effect due to the change in the density of state is rapidly hindered by the effect of the increase of the percolation threshold in contrast to the 3D case. 

\section{Comparison to experiments}

Experimentally, the appearance of coherence have been studied in ultra-cold 2D rubidium disordered gases \cite{Allard2012}. In this experiment, we expect noticeable deviations from the LDA because $\sigma \approx \xi \approx \lambda_\textrm{dB}$. Nevertheless, since there is no quantitative theory in the regime of microscopic disorder explored in the experiment, it is interesting to compare with the LDA approximation results. The two disorder strengths used in the experiment correspond to $V' \approx 0.3$ and $V' \approx 0.7$.  The value $V'=0.7$ is close to the critical disorder and the transition is strongly shifted. This is in agreement with the observation of no sudden appearance of phase coherence in the range of entropy explored in the experiment. The value $V'=0.3$ corresponds to a lowering  of the critical entropy per particle by $0.3\,k_\textrm{B}$. This value is somewhat larger than the shift $\sim 0.2\,k_\textrm{B}$ found in the experiment for the sudden appearance of coherence, a phenomenon linked to the superfluid phase transition. Note that the absolute values of the entropy per particles differ between the above theory and the experiment because of a small fraction of atoms that populates the excited states of the vertical strongly confining harmonic oscillator. We have checked that adding the vertical levels does not significantly affect the value of the entropy shift in the LDA approximation. 

Another quantity that we can calculate in the local density approximation is the change of temperature due to the addition of the disorder. For the parameters of the experiment, the temperature changes are found to be 5\,nK and 14\,nK. These values clearly exceed the experimentally observed heating. The fact that the LDA approximation overestimates the effects of the disorder can be interpreted as a consequence of an effective smoothing of the disorder for $\sigma \approx \xi \approx \lambda_\textrm{dB}$ and thus as a proof of beyond LDA physics. 

\section{Conclusion}

In this paper, we have analyzed 2D and 3D Bose gases in the presence of disorder, which is treated in the local density approximation. The superfluid part of the phase diagram is found as a function of disorder and temperature and as a function of disorder and entropy. Our results show the competition between two effects when adding disorder. One is the appearance of a percolation threshold, which tends to hinder superfluidity. The second is the disorder induced compression of the gas and the change of the density of state. On the contrary, it is favorable for the appearance of superfluidity. Since the percolation thresholds greatly differ in 2D and 3D, we find very different quantitative behaviors of the superfluid transition curves. Although it might be counterintuitive, the addition of a low amount of disorder always lowers the critical temperature as well as the critical entropy. Finally, this works permit quantitative comparisons with experimental findings in the presence of disorder.  

\section{Acknowledgements}

We thank B. Allard, L. Fouch\'{e}, M. Holzmann, T. Plisson, G. Salomon, and L. Sanchez-Palencia for discussions. This research was supported by RTRA Triangle de la physique, ANR, I-Sense. LCFIO is member of IFRAF.




\end{document}